\documentclass[preprintnumbers,amsmath,amssymb,twocolumn]{revtex4}
\usepackage[colorlinks=true,citecolor=red,filecolor=green,linkcolor=blue,pdfnewwindow=true]{hyperref}
\usepackage{graphicx,ulem}
\usepackage{makeidx}
\usepackage{mathrsfs}
\usepackage[caption=false]{subfig}
\usepackage[dvipsnames]{xcolor}
\usepackage{animate}
\usepackage{lipsum}  
\usepackage[title]{appendix}

\newcommand{\di}{\mathrm{i}}
\def\rmp#1#2#3{{ Rev. Mod. Phys.} {\bf #1}, #2 (#3)} 
\def\prl#1#2#3{{ Phys. Rev. Lett.} {\bf #1}, #2 (#3)} 
\def\prr#1#2#3{{ Phys. Rev. Research} {\bf #1}, #2 (#3)} 
\def\pla#1#2#3{Phys. Lett. A {\bf #1}, #2 (#3)}

\def\pre#1#2#3{Phys. Rev. E {\bf #1}, #2 (#3)}

\def\physd#1#2#3{Physica D {\bf #1}, #2 (#3)}

\def\jpa#1#2#3{J. Phys. A {\bf #1}, #2 (#3)}
\def\jsp#1#2#3{J. Stat. Phys. {\bf #1}, #2 (#3)}

\def\physd#1#2#3{Physica D {\bf #1}, #2 (#3)}

\def\cns#1#2#3{Commun. Nonlinear Sci. Numer. Simul. {\bf #1}, #2 (#3)}

\def\chaos#1#2#3{Chaos {\bf #1}, #2 (#3)}
\def\epl#1#2#3{Europhys. Lett. {\bf #1}, #2 (#3)}
\def\screp#1#2#3{Sci. Rep. {\bf #1}, #2 (#3)}

\def\ie{i.e. }

\def\beqr{\begin{eqnarray}}
\def\eqnr{\end{eqnarray}}
\def\beq{\begin{equation}}
\def\bc{\begin{center}}
\def\ec{\end{center}}
\def\eqn{\end{equation}\noindent}
\begin{document}
\title{Effect of noise on explosive synchronization}
\author{ Ruby Varshney, and Haider Hasan Jafri}
\affiliation{Department of Physics, Aligarh Muslim University, Aligarh 202 002, India}
\begin{abstract}
In this paper we explore the emergence of explosive synchronization (ES) in a star network by considering the dynamics of coupled phase oscillators in the presence of noise. While ES has been the subject of many recent studies, in most cases deterministic dynamics was considered to explore the first-order phase transition. This raises the issue of how robust ES is in situations where fluctuations cannot be suppressed. Thus, to address this issue, we consider a situation where the natural frequency is considered to be correlated with their degrees. We observe that noise plays a crucial role when it is present in the hub.  By considering the model examples of Kuramoto and Stuart Landau oscillators on each node, we examine the effect of noise strength in the hub.
\end{abstract}
\maketitle

% *********************Section 1******************************

\begin{section}{Introduction}
\label{sec1}
%%%%%%%%%%%%%%%%%%%%%%%%%%%%%%% ES  %%%%%%%%%%%%%%%%%%%%%%%%%%%%%%%%%%%%%%%%%%%%%%%%%%%%%%%%%%
Network of coupled oscillators has proven to be a useful paradigm for understanding diverse processes in various fields, particularly physics, biology, engineering and neuroscience. 
In the past few years, there has been a resurgence of interest in exploring the emergent dynamics in networks of coupled oscillators to understand various dynamical behaviors, namely clustering, pattern formation, synchronization \cite{Pikovsky-Book, Boccaletti-Book}, chimera states \cite{abrams} etc. 
Synchronization of an ensemble of interacting units refer to a transition from an incoherent state to a coherent state and is ubiquitous in nature \cite{Pikovsky-Book, Boccaletti-Book, Strogatz-Book}.
 It has been demonstrated that the network topology plays an important role in the emergence of synchronization \cite{Arenas-2006, Barahona-2002,Arenas-2008, watts}.
Most studies have reported that the transition to synchrony is continuous in nature and is a second order transition \cite{ Gardenes-2007, Nishikawa-2003, Kuramoto-Book, Acebron-2005}. 
However, in scale free (SF) networks, a discontinuous transition to synchrony termed as explosive synchronization (ES) \cite{Gardenes-2011} has attracted attention of many researchers.

 ES  has two important features: a discontinuous transition and a hysteresis associated with the backward and forward transitions. This discontinuous or first-order transition is considered to be an outcome of a positive correlation between the node degree and the corresponding oscillator’s natural frequency. This transition has been studied extensively in phase oscillators \cite{ Gardenes-2011, Peron-2012, Zhang-2013, Hu-2014, Zhou-2015, Coutinho-2013, Zou-2014, Vlasov-2015}. In Ref.~\cite{Zhou-2015}, it is shown that the nature of transition changes from first-order to second-order as the central frequency of the frequency distribution shifts in the positive direction. Recently, it has been observed that ES can occur as a result of positive correlations between the coupling strengths of the oscillators and the absolute of their natural frequencies \cite{Zhang-2013} and time delay \cite{Yeung1999}. Apart from phase oscillators, first order transitions to synchrony have been observed in cases of oscillators having more complicated dynamics, namely limit cycle oscillators \cite{ Bi-2014}, chaotic oscillators \cite{Levya-2012} and excitable systems \cite{Chen-2013, Boaretto-2019}. Effect of various frequency distributions on the nature of phase transitions has been explored in Refs.~\cite{Hu-2014, Zhou-2015, Bi-2014}. It has been reported that inclusion of inertia terms in second order Kuramoto oscillators give rise to a discontinuous synchronization transition, namely cluster ES  where nodes join the synchronous components in the form of clusters \cite{Ji-2013, Ji-2014}.  Emergence of ES has been observed by including an adaptive factor from the global order parameter in the coupling term \cite{Filatrella-2007}. Considering multiple connections for a given node, the notion of ES has been explored in multiplex networks \cite{Jalan-2019, Kumar-2020, Kumar-2021}. 
 
 %%%%%%%%%%%%%%%%%%%%%%%%%%%%%NOISE%%%%%%%%%%%%%%%%%%%%%%%%%%%%%%%%%%%%%%%%%%%%%%%%%%
 Naturally occurring systems are subjected to random external fluctuations or noise \cite{Chen-2004, Greenman-2003, Erguler-2008, Surovyatkina-2005, Alonso-2007, Ojalvo-1996}. In certain situations where noise is inherent, the fluctuations may induce undesired states. In other situations, noise can play a constructive role. It can induce stochastic resonance in the system which helps in detection of weak-signals \cite{Benzi-1981,Longtin-1998}. In case of multistable systems, noise can make the system to hop between its attractors \cite{Kraut-1999,Kaneko-1997,Zerega-2012,Pisarchik-2009}, thereby controlling the multistability.
 
 The study of synchronization of chaotic systems in the presence of noise has attracted much attention and is found to occur widely in nature \cite{Glass-2001}. Example ranges from the case of coupled weather system to neurons or chemical to chaotic oscillators \cite{Lloyd-1999,Neiman-1999,Tavazoie-1999, Tu-2005}.  
 There are situations where single noise process referred to as common noise may influence the entire system. This can actually induce order, and has been studied extensively for an ensemble of periodic oscillators \cite{Uchida-2004, Nagai-2010}.  
 However, there are many situations where individual oscillators may evolve under the influence of independent noise resulting in the  inhibition of synchrony \cite{Sakaguchi-1998}.  
Most studies have focused on the influence of either common noise or intrinsic noise without any degree-frequency correlation. 
However, it has recently been reported that under the influence of noise and stochastic perturbations, the synchronization changes from cluster explosive synchronization (CES) to non-CES \cite{Cao-2018} for smaller degree nodes. 

In the present work, we investigate the behavior of phase transition in a star network where the frequency of the oscillator is positively correlated to its degree.
The star network  consists of a central hub  connected to $N$ nodes.
Since all the nodes are connected to a common hub, they evolve under the influence of a common noisy signal.
Thus, in this work we simultaneously explore the two different origins of cooperative behavior emerging in distributed systems: coupling and common noise. 
We study the phase transition by considering the dynamics governed by the Kuramoto oscillators in the presence of noise. For this case, we calculate the transition points using semi-analytical tools. The study is further extended to validate the findings in case of the Stuart-Landau oscillators with common noise.
The paper is organized in the following manner. In Section~\ref{sec2}, we study the ES on a star network where the dynamics on each node is that of a Kuramoto oscillator. We discuss the dynamical scenarios that are observed as a result of introducing noise in the hub. This study is further extended to substantiate our results by considering the dynamics of Stuart-Landau oscillators in Section~\ref{sec3}. A summary of our findings are presented in Section~\ref{sec4}.

 \end{section}
%************************Section 2**********************************

\section{Stochastic Kuramoto oscillators}
\label{sec2}

We consider a star network which consists of a central hub connected to $N=500$ nodes. The dynamics on top of each node is that of the Kuramoto oscillator. 
To incorporate the effect of noise, we consider stochastic first order Kuramoto oscillators, for which the equations of motion are given by 
\beqr
	\label{4}
 \dot{\phi}_j &=& \omega+\lambda\sin(\psi-\phi_j),    \nonumber\\
 \frac{1}{\beta}\dot{\psi} &= &\omega+\frac{\lambda}{N} \sum_{j=1}^{N} \sin(\phi_j-\psi)+ \eta D_H,
\eqnr
where $j = 1,2 \dots N$, $N$ being the total number of nodes in the system, $\phi_j$ is the phase of the oscillator at the $j$-th node, $\psi$ is the phase of the oscillator at the hub, $\omega$ is the frequency of all the oscillators, $\beta$ is the parameter for frequency mismatch, $\lambda$ is the strength of coupling between the nodes and the hub and $D_H$ is the strength of noise in the hub. $\eta$ is the $\delta-$correlated Gaussian white noise that satisfies, $\langle\eta(t)\rangle=0$  and $\langle\eta_k(t)\eta_l(t')\rangle=\delta(t-t')\delta_{kl}$.

To describe the degree of coherence in the network, we define the global order parameter $R$ as
\begin{equation}
\label{2}
R = \frac{1}{N} \left| {\sum_{j=1}^{N} e^{i(\phi_j-\psi)}} \right|.
\end{equation}

 We have used the standard RK4 method with a step size of $10^{-3}$ 
 to calculate the order parameter $R$ for forward and backward transitions.
  For forward transition, we consider uniformly distributed phases in the interval [0,2$\pi$] as initial conditions and then the final state of the previous $\lambda$ is taken as the initial state for the next $\lambda$. 
 In the backward direction, for each $\lambda$, we start with initial conditions that are close to the synchronized state and the coupling strength is varied adiabatically by an amount $\Delta \lambda= 0.05$.

\begin{figure}[htp]
\includegraphics[scale=0.6,angle=270]{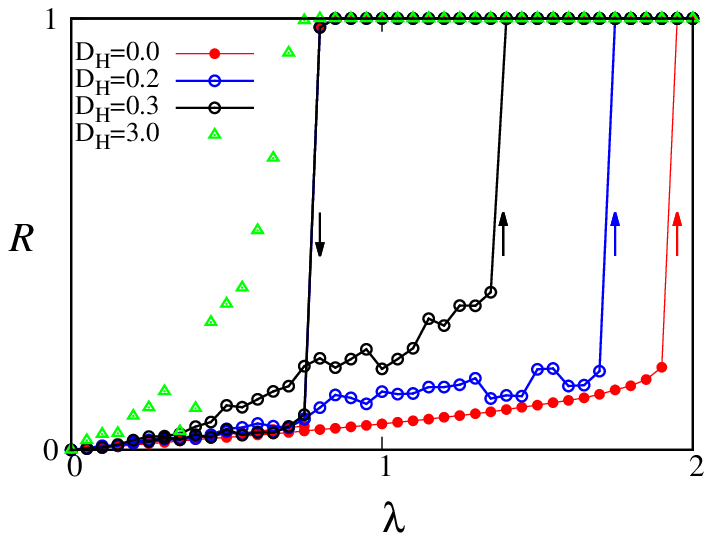}
\caption{Explosive synchronization with forward and backward transitions in a star-network of $N=500$ nodes and a hub with $\beta=10, \omega=1$ represented by Eq.~\eqref{4} for different values of $D_H=0,0.2,0.3$. A second order transition is observed for higher values of noise strengths in hub ($D_H=3.0$, green triangles).}
\label{fig1}
\end{figure}
  In the present work, the effect of noise in the system is studied systematically when noise is present in the hub.
We observe that both the coupling strength and the noise strength have a notable influence on the nature of transition to synchrony.
We comprehend the situation by plotting the order parameter $R$ shown in Fig.~\ref{fig1}.
  Here, we show the dependence of the order parameter $R$ on $\lambda$ for $D_H=0, 0.2$ and $0.3$ and observe that $R$ abruptly jumps from  $R \approx 0$ to $R \approx 1$ for the forward transitions and vice versa in case of backward transition.
  This behavior of order parameter $R$ is a signature of explosive first order transition. 
We also notice that as the value of $D_H$ increases, the nature of first order transition is still present but the area of the  hysteresis loop decreases.  
It may be noted that the forward transition point %($\lambda_c^f$)
shifts to a new position whereas the backward transition does not change and hence the hysteresis area changes.
For higher values of $D_H$ (say $D_H=3$), the system makes a second order transition to synchrony as shown in Fig.~\ref{fig1}(green triangles). 

From the above discussion, we conclude that the presence of noise in the hub plays an important role. In a star network, if the noise in present in the hub (common noise), then not only the transition points but also the nature of the transition can be changed. This is due to the fact that hub has higher degree and is coupled to all other nodes. 
 \begin{figure}[htp]
\includegraphics[scale=0.6,angle=270]{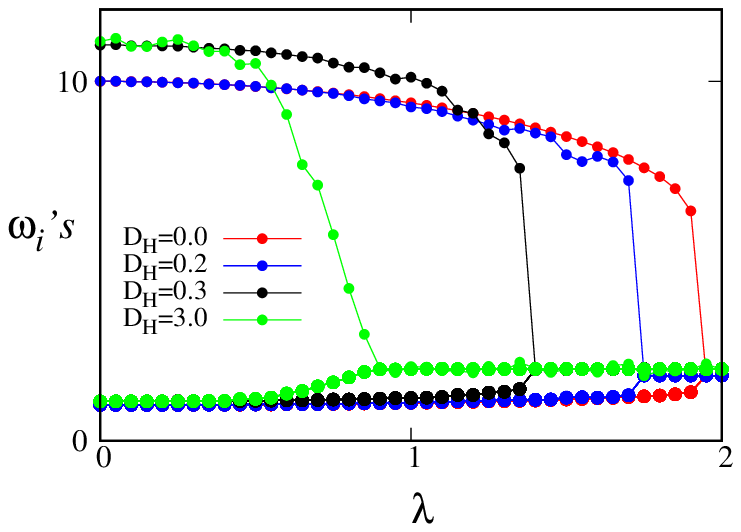}
\caption{The frequency of the oscillators in the star network (Eq.~\eqref{4}) along the forward direction for four different values of noise strengths $D_H$. Frequencies of all the nodes join the major synchronous component at the forwards transition point. Note that with increasing $D_H$, transition to synchrony is achieved for smaller values of $\lambda$.}
\label{fig2}
\end{figure}

\subsection{Frequency and Phase plots}

 To analyze the dynamics at the microscopic level, we calculate the effective frequencies of individual oscillators by using
\beq
\omega_i^{\text{eff}}=\frac{1}{T}\int_t^{t+T}\dot{\theta}_i(t)d \tau ,
\eqn
where $T$ is the total time.
The variation of frequencies of the oscillators with the coupling strength $\lambda$, shown in Fig.~\ref{fig2}, clearly describes how the synchronization state is achieved. 
We plot frequencies for the cases considered in plotting Fig.~\ref{fig1} which include both the first and second order transition to synchrony. 
At $\lambda=0$, the frequency of hub is close to $\omega_h=10$ while the frequency of all the nodes are $\omega_i =1$ at all $D_H$ values.  Ultimately, all the frequencies merge at the forward transition point. 
We note that all frequencies join the major synchronous component through an abrupt transition for $D_H=0,0.2$ and $0.3$ while it's a continuous transition to synchrony for $D_H=3.0$.
\begin{figure}[ht]
\subfloat[]{\includegraphics[width = 45mm,height=45mm]{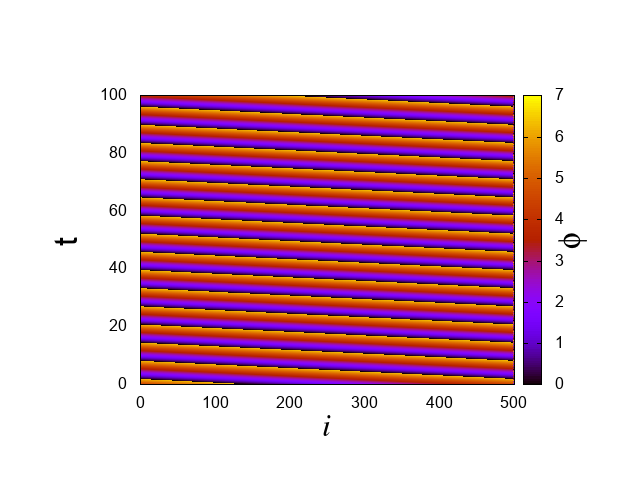}} 
\subfloat[]{\includegraphics[width = 45mm,height=45mm]{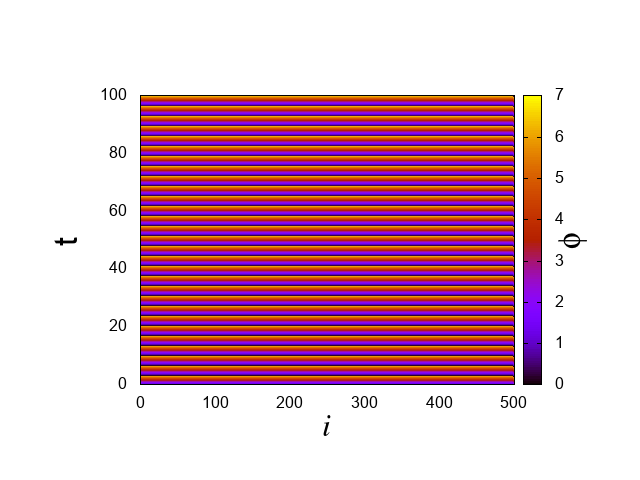}} \\
\subfloat[]{\includegraphics[width = 47mm,height=40mm]{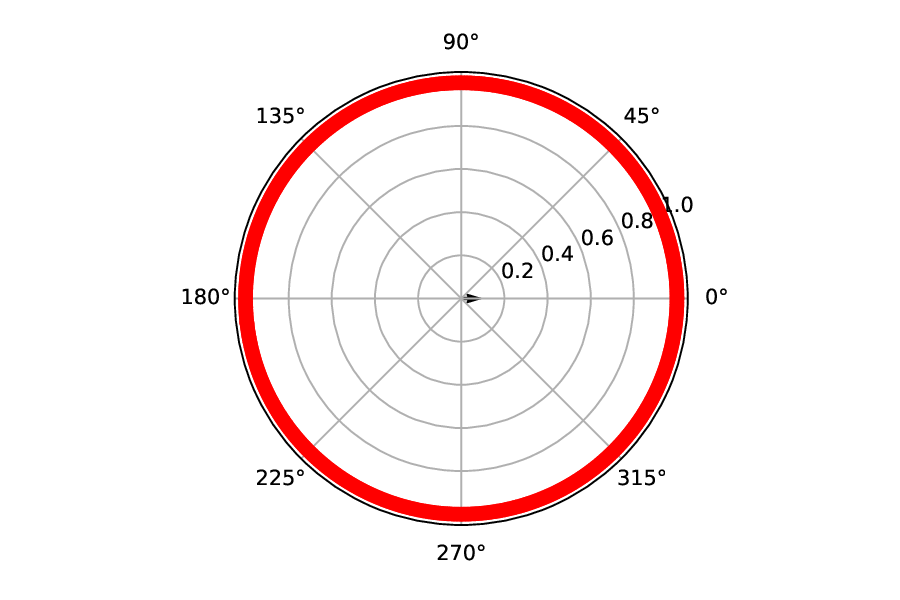}}
\subfloat[]{\includegraphics[width = 47mm,height=40mm]{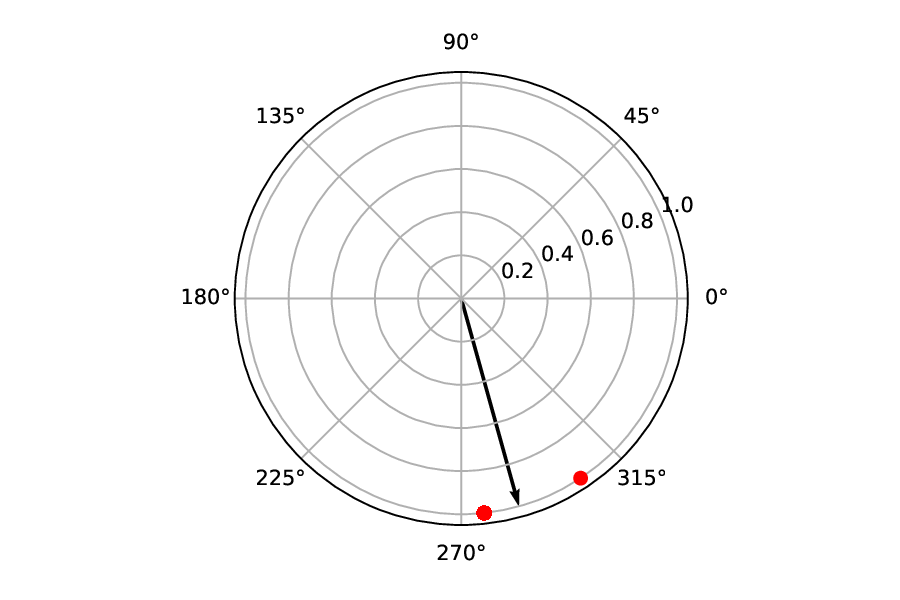}}
\caption{Variation of phase of the $i$-th oscillator in time and space for the Kuramoto oscillators at $D_H=0.2$. The phase-time plot is shown for (a) $\lambda=0.05$ and (b) $\lambda=1.75$. Similarly, the polar plot is shown for (c) $\lambda=0.05$  and (d) $\lambda=1.75$. Solid black arrow shows the magnitude of the order parameter $(R)$.}
\label{fig3}
\end{figure}

\begin{figure}[]
\includegraphics[width = 65mm,height=95mm,angle = 270]{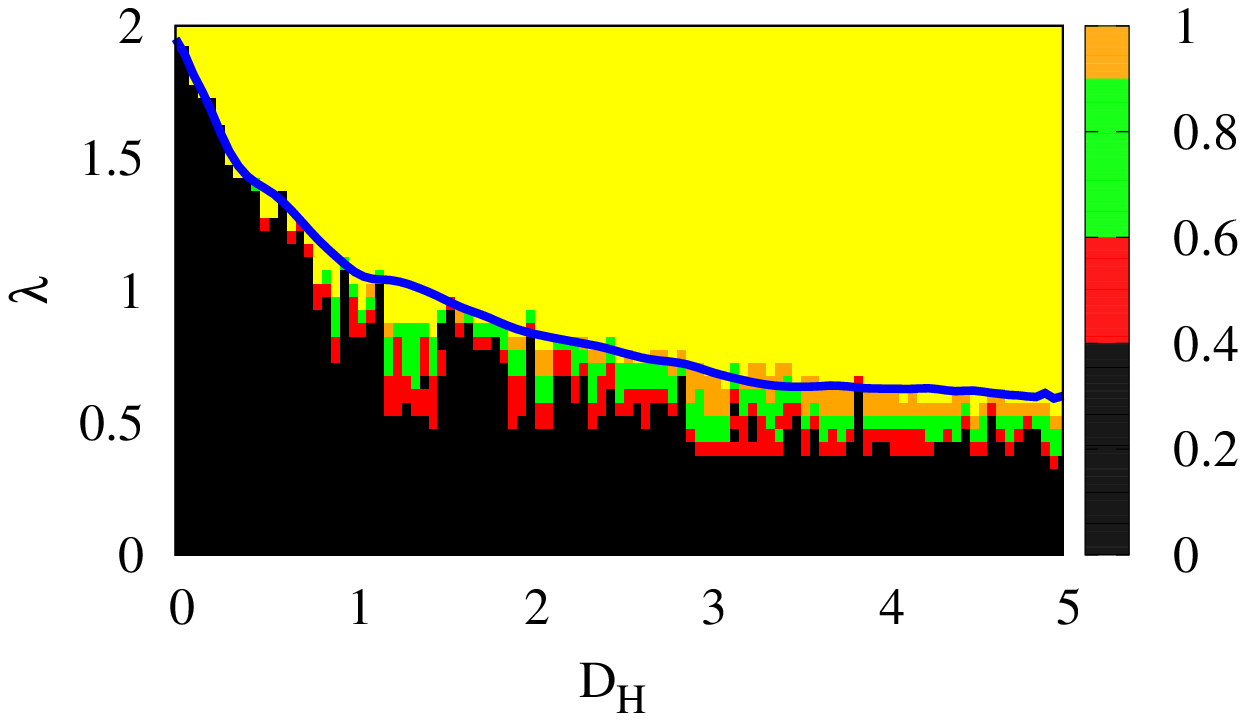}
\caption{Parameter space diagram in $D_H-\lambda-R$ space with $D_L=0$. Color axis shows the value of the order parameter (R). Blue line represents the curve for phase synchrony obtained by solving Eq.(\ref{14}) for $\rho$-parameter in the semi-analytical approach.}
\label{fig4}
\end{figure}

This can further be seen in the phase-time plots and radar representation of phases of the system at $D_H=0.2$ as shown in Fig.~\ref{fig3}. 
For lower values of the coupling strength $(\lambda=0.05)$ the phases are not synchronized as shown in the phase-time plot (Fig.~\ref{fig3}(a)). 
Similarly, at this coupling value, we note that the phases rotate with different frequencies and all the phases are well distributed on a unit circle, making the order parameter to remain close to zero, as shown in Fig.~\ref{fig3}(c). 
At a higher value of coupling strength  $(\lambda=1.75)$, the phases are locked in time as shown in Fig.~\ref{fig3}(b) and rotate with same frequency as described in Fig.~\ref{fig3}(d).

\subsection{Parameter space}
We explore the parameter space by calculating the order parameter $R$. We plot the order parameter $R$ in $\lambda-D_H$ plane as shown in Fig.~\ref{fig4}.
 To understand the effect of coupling strength in the presence of noise, we first fix the value of $D_H$ and study the variation of $R$ with coupling parameter $\lambda$. We repeat this process with a change in the $D_H$-value. 
 If the noise strength is small, we observe that the transition from desynchrony (black) to synchrony (yellow) is first order in nature. 
 However, if we increase the noise strength, synchronization is achieved even for smaller values of $\lambda$ but the order parameter would change continuously, indicating that the transition is second order in nature. 
 Thus, we note that synchrony may be achieved even for small values of $\lambda$ if the noise is present in the hub.
Further, we observe that, for higher $D_H$ values, noise induces phase synchrony even in the region below the backward transition point.

\subsection{Mean field analysis}
\label{mfa}
We discuss a semi-analytical approach to identify the region of synchrony in stochastic Kuramoto oscillators where all the nodes are identical. Following the Watanabe-Strogatz (WS) approach given in ref.~\cite{Vlasov-2015}, we solve our model in terms of the order parameter by defining a phase difference,
\beq
\Phi_j = \phi_j - \psi .
\eqn
Thus, the system Eq.(\ref{4}) can be rewritten as,
\beq
\dot{\Phi}_j = -(\beta-1)\omega-\beta D_H \eta - \frac{\beta\lambda}{N} \sum_{j=1}^{N} \sin(\phi_j-\psi) + \lambda \sin(\psi-\phi_j),
\eqn
which reduces to a form,
\beq
\label{11}
\dot{\Phi}_j = -(\beta-1)\omega-\beta D_H \eta - \beta\lambda \operatorname{Im}(H(t)) + \lambda \operatorname{Im}(e^{-\di\Phi_j}), 
\eqn
with
\beq
H(t) = \frac{1}{N} \sum_{j=1}^{N} (e^{\di \Phi_j}), \nonumber
\eqn
where $j=1,2 \dots N$. This N-dimensional equation can be reduced to lower dimensional equation using WS ansatz \cite{Watanabe-1993,Watanabe-1994} for the general form
\beq
\label{12}
\dot{\theta_i} = g(t) + \operatorname{Im}(G(t) e^{-\di \theta_i}) .
\eqn
 By comparing Eq.(\ref{11}) with Eq.(\ref{12}), we get
  \beqr
  g(t) &=& -(\beta-1)\omega-\beta D_H \eta - \beta\lambda \operatorname{Im}(H(t)), \nonumber \\
  G(t) &=& \lambda.
  \eqnr
We transform N variables $\theta_i$ of Eq.(\ref{12}) to a set of global variables $z,\alpha$ ($z$ being a complex and $\alpha$ being a real variable) using the M\"{o}bius transformation \cite{Marvel-2009,Pikovsky-2015}
\beqr
e^{\di \theta_i} = \frac{z+e^{\di(\zeta_i + \alpha)}}{1+z^{*} e^{\di(\zeta_i + \alpha)}} \quad,
\eqnr
where $\zeta_i$ are additional constraints. In the transformed set of variables, Eq.(\ref{12}) is written as
\beqr
\dot{z} &=& \di g(t) z + \frac{G(t)}{2}-\frac{G(t)^*}{2} z^2, \nonumber \\
\dot{\alpha} &=& g(t) + \operatorname{Im} (z^* G(t)) .
\eqnr
Now, we can re-write the equations of motion for the z-variable,\beqr
\label{14}
\dot{\rho} &=& \frac{\lambda}{2} (1-\rho^2) \cos(\varphi), \nonumber \\
\dot{\varphi} &=& -(\beta-1)\omega-\beta D_H \eta - \beta\lambda \rho \sin(\varphi) \nonumber\\
&&- \frac{\lambda}{2\rho}\sin(\varphi)(1+\rho^2), 
\eqnr
where $z = \rho e^{\di \varphi}$, $\rho$ is the order parameter. Now we numerically integrate the coupled equations Eq.(\ref{14}) to obtain the values of $\rho$ as a function of $\lambda$ and  $D_H$. The result, thus obtained, is plotted in Fig.~\ref{fig4} and is shown by blue curve. This blue curve separates the region of phase synchrony from the rest of the space. Thus, the result of semi-analytical approach matches well with the numerical results.

% ************************Section 3*********************************

\section{Stuart-Landau oscillators}
\label{sec3}
In a number of situations of practical importance, it is of interest to examine how the ideas of ES can be extended to a system with amplitude dynamics where fluctuations cannot be suppressed. As an example of the scenario, we consider a star network of Stuart-Landau oscillators with frequency-weighted coupling represented as %\\[-1em]

\beqr
\label{5}
\dot{z_j}(t) & = &(1+i\omega_j-|z_j|^{2})z_j(t) +\lambda |\omega_j|(z_h(t)-z_j(t)),\nonumber\\
\frac{1}{\beta}\dot{z_h}(t) & =&(1+i\omega_h-|z_h|^{2})z_h(t)  + \frac{\lambda |\omega_h|}{N} \sum_{j=1}^{N}(z_j(t)-z_h(t)) \nonumber \\
\eqnr
where $z_j(t)$ represents the complex amplitude of oscillator on the $j$th-node ($j=1,2,\dots,N$), $N$ being the total number of nodes in the star ($N=500$) and $z_h(t)$ represents the complex amplitude of the oscillator on the hub. $\beta$ is the scale separation parameter and $\lambda$ is the coupling strength. In Ref.~\cite{Bi-2014}, a system of frequency-weighted globally coupled SL oscillator has been used to observe explosive death. In our work, we consider frequency weighted coupling along with degree frequency correlation in a star network. Frequency of the nodes ($\omega_j$'s) are drawn from two different frequency distributions (FDs), namely the triangular distribution expressed as \cite{beta},
\[   
f(\omega) = 
     \begin{cases}
       4 \omega ,\hspace{12mm} \text{for} \quad 0 \leq \omega < \frac{1}{2},   \\
       4 (1-\omega) ,\quad \text{for} \quad \frac{1}{2} \le \omega \leq 1, \\
       0, \qquad  \qquad  \text{elsewhere} 
     \end{cases}
\]
and the uniform distribution given by,
 \[   
f(\omega) = 
     \begin{cases}
      1 ,\hspace{4mm} \text{for} \quad 0 \leq \omega \leq 1,   \\
      0,\quad \text{for} \quad  \omega < 0 \quad \text{or} \quad \omega > 1.
     \end{cases}
\]

Transition to synchrony is observed using both the amplitude order parameter $R_{amp}$  and the phase order parameter $R_{phase}$ which are defined as
\begin{equation}
R_{amp} = \frac{1}{N} \left| {\sum_{j=1}^{N} z_j(t)} \right|, \quad
R_{phase} = \frac{1}{N} \left| {\sum_{j=1}^{N} e^{i(\phi_j)}} \right| ,
\end{equation}
where $\phi_j = \tan^{-1}(y_j/x_j)$ is the phase, $y_j$ and $x_j$ being the real and imaginary component of the complex amplitude $z_j$. Thus, with the help of these two order parameters it is convenient to characterize the dynamics of both the amplitude and the phase.

In presence of noise, the dynamics of the SL oscillators on a star network may be mathematically described by rewriting Eqs.~\eqref{5} as follows
\beqr
\label{6}
\dot{z}_j(t) & =& (1+i\omega_j-|z_j|^{2})z_j(t)  \nonumber\\
&+&\lambda |\omega_j|(z_h(t)-z_j(t)),\nonumber \\
\frac{1}{\beta}\dot{z}_h(t) & =& (1+i\omega_h-|z_h|^{2})z_h(t) \nonumber\\
&+& \frac{\lambda |\omega_h|}{N} \sum_{j=1}^{N}(z_j(t)-z_h(t)) + {\eta D_H} ,
\eqnr
where $\eta$ is the $\delta-$correlated Gaussian white noise introduced in Sec.~\ref{sec2} and $D_H$ is the noise strength in the hub. 

\begin{figure} 
\includegraphics [scale=0.4,angle=270]{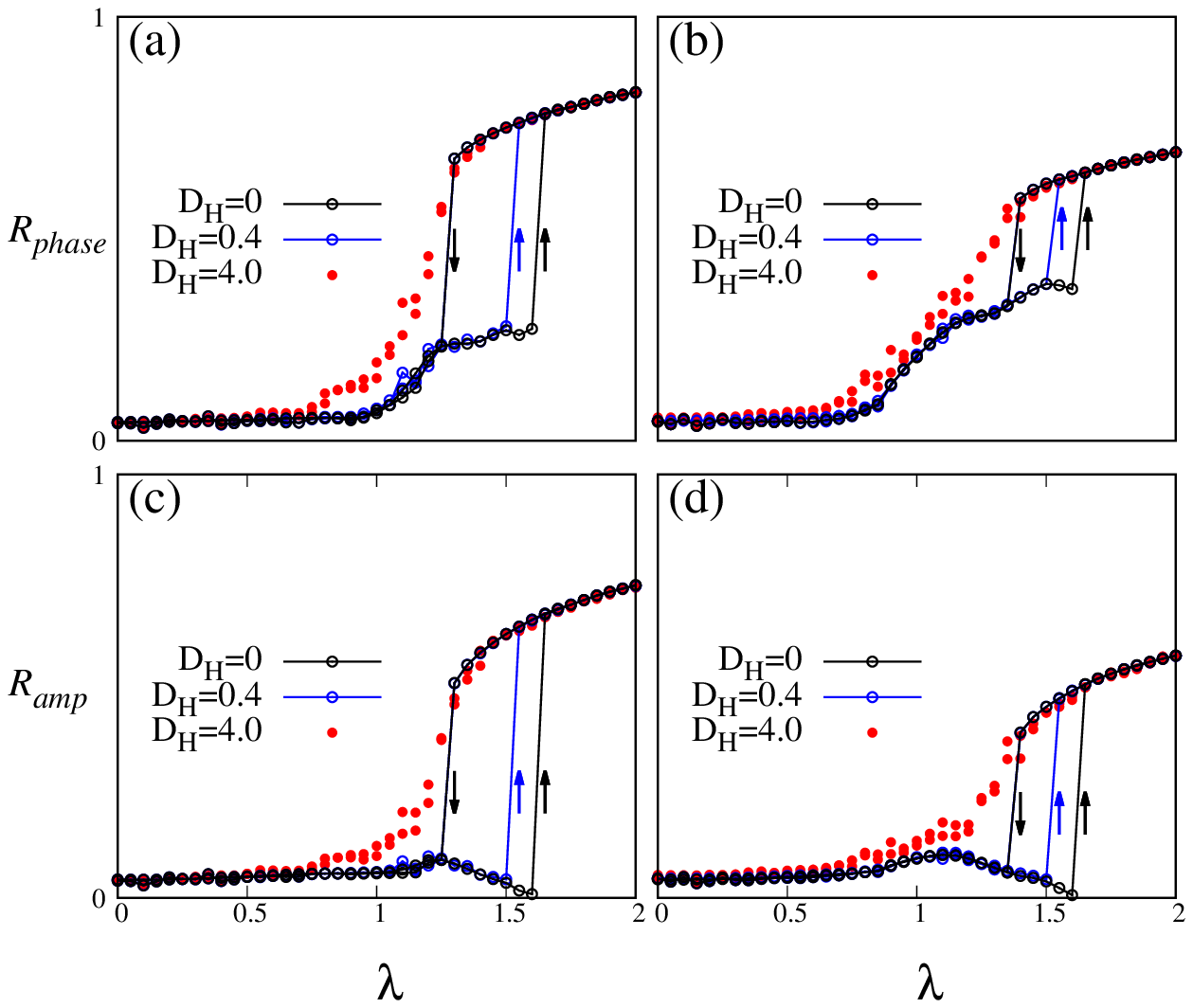}
\caption{Variation of phase order parameter $R_{phase}$ and amplitude order parameter $R_{amp}$ with coupling strength ($\lambda$) for two types of FDs at three different values of noise strength in hub $D_H=0, 0.4$ and $4.0$. Upward and downward arrows show the forward and backward transitions respectively indicating an explosive synchronization in a star-network (Eq.~\eqref{6}). $R_{phase}$ and $R_{amp}$ for triangular FD are plotted in (a) and (c) respectively while for uniform FD these are plotted in (b) and (d) respectively.}
\label{fig5}
\end{figure}

The basic framework of our study explores the effect of $D_H$ on collective behavior.
 When noise is present in the hub, we plot the order parameters ($R_{amp}$ and $R_{phase}$) with coupling strength ($\lambda$) for the two frequency distributions (FDs) as shown in Fig.~\ref{fig5}. 
When the frequencies are drawn from a triangular distribution, the variation of the order parameters $R_{phase}$ and $R_{amp}$ are shown in Figs.~\ref{fig5}(a) and \ref{fig5}(c) respectively. 
 We observe that for $D_H=0$  (black circles) both the order parameters show a discontinuous transition from an incoherent state to the coherent state and vice versa in both the forward and the backward continuations. 
 Further, this transition is also accompanied by a well defined hysteresis. If the noise strength in the hub is increased, \ie at $D_H=0.4$, the hysteresis width decreases as shown by the blue circles.    
 Finally, at large value of $D_H$ ($D_H=4.0$), the hysteresis disappears and the transition becomes a second order transition as depicted by the red dots in Figs.~\ref{fig5}(a) and \ref{fig5}(c).

 In case of the uniform FD, we find qualitatively similar results as that of triangular case. In Figs.~\ref{fig5}(b) and \ref{fig5}(d), we plot the order parameters $R_{amp}$ and $R_{phase}$ respectively with the coupling strength $\lambda$.
  In the absence of the noise $D_H=0$, we observe a discontinuous transition in both the forward and backward continuations followed by a hysteresis.
  This is shown by the black circle curve in Figs.~\ref{fig5}(b) and \ref{fig5}(d). As the noise strength increases, we observe a  decrease in the hysteresis width for $D_H=0.4$ as described by blue circle curve in the same figure. 
  Finally, for large noise strength in the hub $(D_H=4.0)$, the hysteresis vanishes and the transition becomes a second order transition (shown in red color dots in Figs.~\ref{fig5}(b) and \ref{fig5}(d)).
  
Thus, we observe that for both the FDs, the hysteresis width decreases with increase in the noise strength in the hub ($D_H$) and finally for large values of $D_H$, we observe that the transition becomes second order in nature and the hysteresis vanishes.

 Similar observations have been made if we consider unweighted coupling in the system Eqs.~\eqref{5}  except for a smaller hysteresis.

\begin{figure}[htp]
\includegraphics [scale=0.4,angle=270]{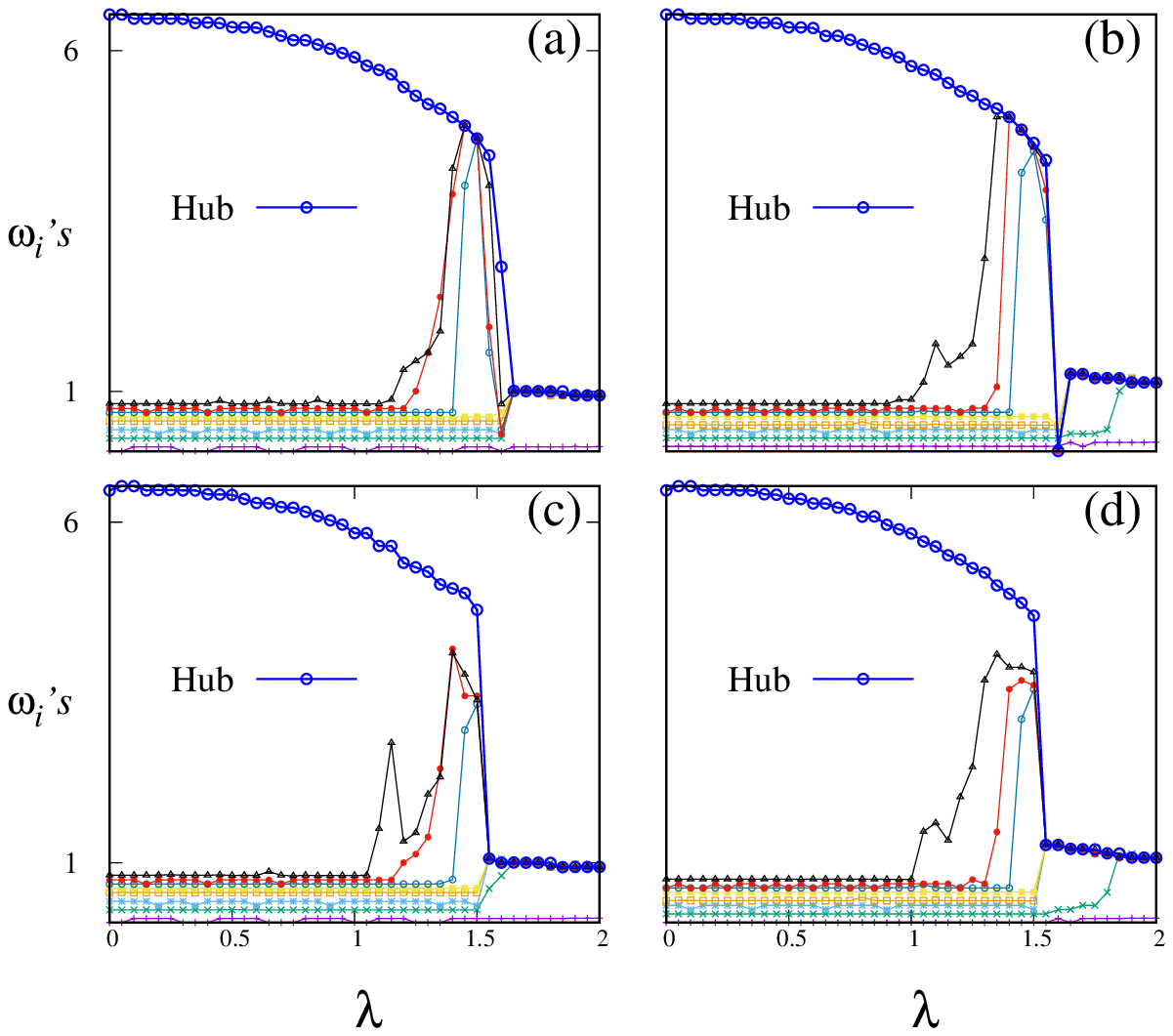}
\caption{Frequency variation of 10 oscillators in the star network (Eq.~\ref{6}) with the coupling strength $\lambda$ at  $D_H=0$ for (a) triangular FD and (b) uniform FD. At $D_H=0.4$ the frequency variation is shown for (c) triangular FD and (d) uniform FD. Note that in each case we consider a forward increment in $\lambda$.}
\label{fig6}
\end{figure}
 
\subsection{Frequency plots and Phase trajectories}
To investigate further the underlying dynamics behind the ES witnessed in the star network of coupled Stuart-Landau oscillators we perform a detailed study of the dynamics of individual node. Thus, we explore the emergent dynamics of the system by plotting the variation of frequency of the oscillators with the coupling strength $\lambda$. To calculate frequency in presence of noise we make use of Hilbert phase as described in Ref.~\cite{Rosenblum-1996}. Given a signal $s(t)$, define the Hilbert transform 
\beq
\bar{s}(t)=\frac{1}{\pi}PV \int_{-\infty}^{\infty} \frac{s(\tau)}{t-\tau} d \tau ,
\eqn
where $PV$ denotes the Cauchy principal value. The analytical signal can be constructed as $s(t) + i \bar{s}(t)$ for which the instantaneous amplitude $A(t)$ and phase $\phi(t)$ are related as 
\beq
A(t)e^{i \phi(t)}=s(t) + \bar{s}(t).
\eqn
The instantaneous frequency can be calculated from the slope of the phase growth.
We have used the time series of the real component of the complex amplitude of the oscillator to calculate its frequency. 
We find that initially when the coupling is off, all the oscillators are oscillating with their natural frequencies: all the nodes have frequency $\omega_i \in [0:1]$ while the hub has a larger frequency. %
In the absence of noise, we plot the frequencies of the oscillators with coupling strength. As we increase the coupling strength, most frequencies merge at the forward transition point. 
When the natural frequencies are drawn from the triangular distribution, the variation of the frequencies are shown in Fig.~\ref{fig6}(a) whereas for the uniform distribution, we describe the frequency variations in Fig.~\ref{fig6}(b).
One can see that at the forward transition point, most of the oscillators are synchronized and form a large synchronous component which oscillate with a common frequency explaining the sudden jump in synchrony of the system. 
The leftover nodes gradually join the common frequency curve at some higher value of coupling parameter $\lambda$. This explains the continuous increase in the order parameter in the region after sudden jump.
In presence of noise, the variation of frequencies is plotted in Figs.~\ref{fig6}(c) and \ref{fig6}(d) for triangular and uniform frequency distributions respectively.
We note that most frequencies merge at smaller $\lambda$-values as compared to the noise-free case. Thus, the nature of transition in presence of noise remains the same except the value of forward transition that now occurs for smaller values of $\lambda$. This shift in the forward transition point can also be seen in Fig.~\ref{fig5}. 

\begin{figure}[h]
 \includegraphics[scale=0.4,angle=270]{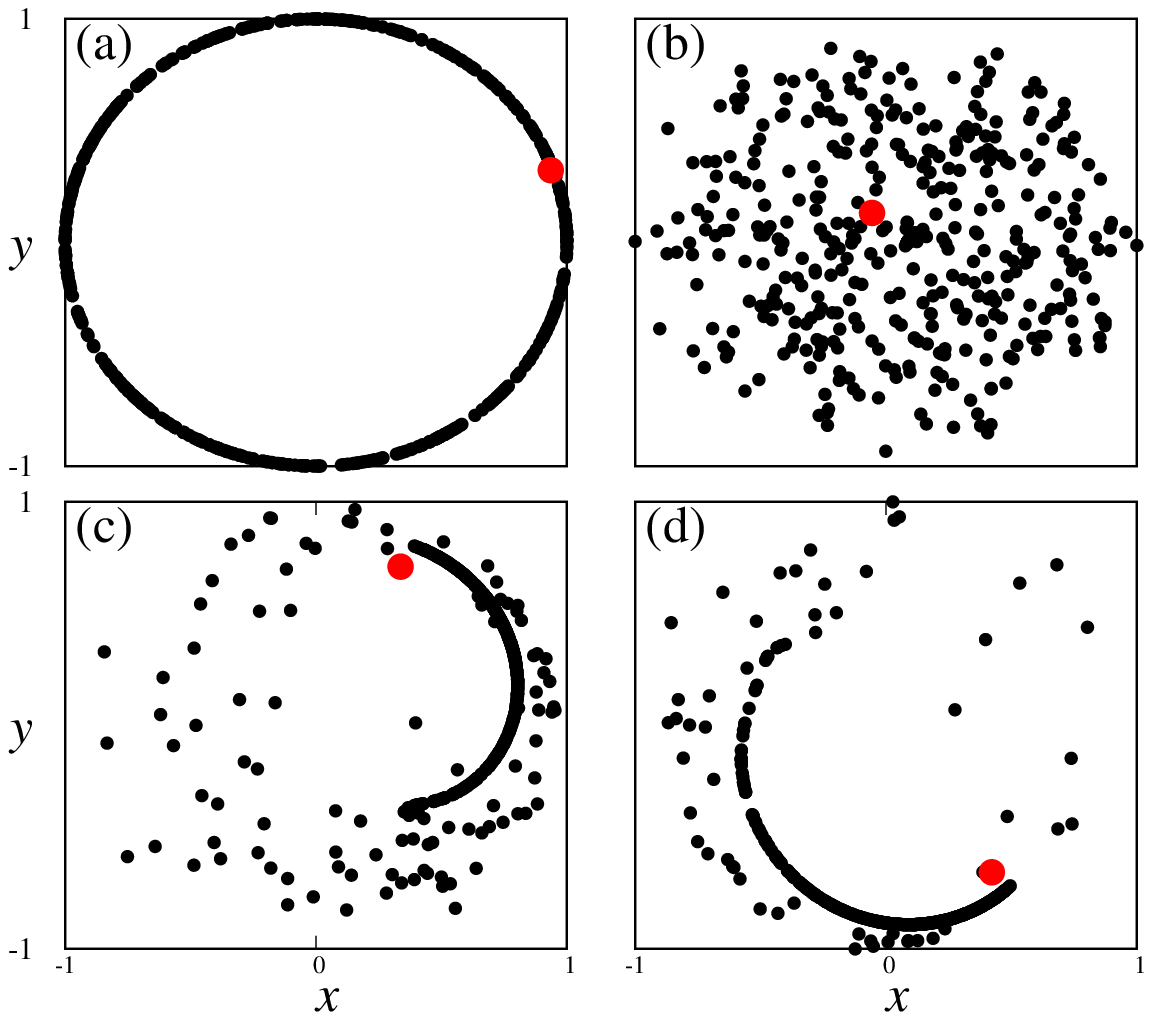}
 \caption{Snapshot of the oscillator's state in the $x-y$ plane taken at sufficiently large time in the presence of noise ($D_H=0.4$) for triangular FD at (a) $\lambda=0.0, R_{amp}=0.04$, (b) $\lambda=1.5,R_{amp}=0.04$, (c) $\lambda=1.55,R_{amp}=0.64$ and (d) $\lambda=2.0,R_{amp}=0.74$. Black dots are for oscillators at nodes and the red dot is for the hub oscillator. Similar observations have been found for uniform distribution.}
 \label{fig7}
 \end{figure}

 In case of Stuart-Landau oscillators where natural frequencies are drawn from a distribution, we observe that  at $\lambda=\lambda_c^f$ (forward critical  transition point), system goes from an incoherent state to a partially coherent state. This happens because higher frequency oscillators are synchronized first while lower frequency oscillators are still drifting around (see Fig.~\ref{fig7}). The leftover nodes are the low frequency oscillators \cite{Zhou-2015, Matthews-1991}. This is shown by the instantaneous states of the oscillators for $D_H=0.4$ in the $x-y$ plane where the frequencies of the oscillators are drawn from the triangular FD. In the absence of coupling $(\lambda=0)$, the phase of the oscillators are uniformly distributed on a circle resulting in a desynchronized state as shown in Fig.~\ref{fig7}(a). When the coupling is just below the transition point, \ie $\lambda=1.5$, the oscillators are still desynchronized as shown in Fig.~\ref{fig7}(b). However, on increasing the coupling further, we observe that the system becomes ordered for $\lambda=1.55$ and $\lambda=2$ as shown by the clustering of the oscillators in Figs.~\ref{fig7}(c) and \ref{fig7}(d).

 \subsection{Parameter space}
\begin{figure}[t]
 \subfloat[]{\includegraphics[height=80mm,width=50mm,angle=270]{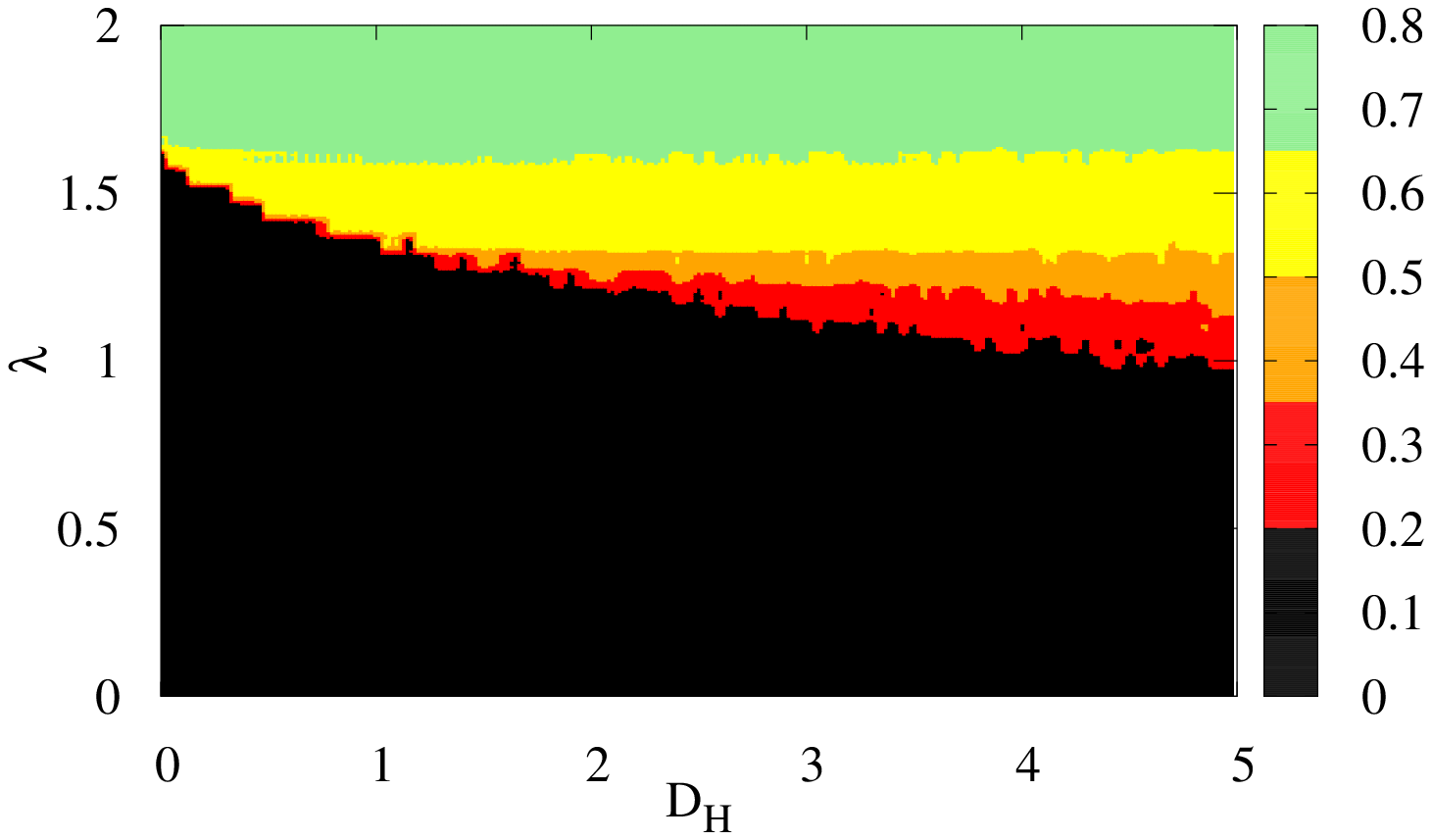}} \
 \subfloat[]{\includegraphics[height=80mm,width=50mm,angle=270]{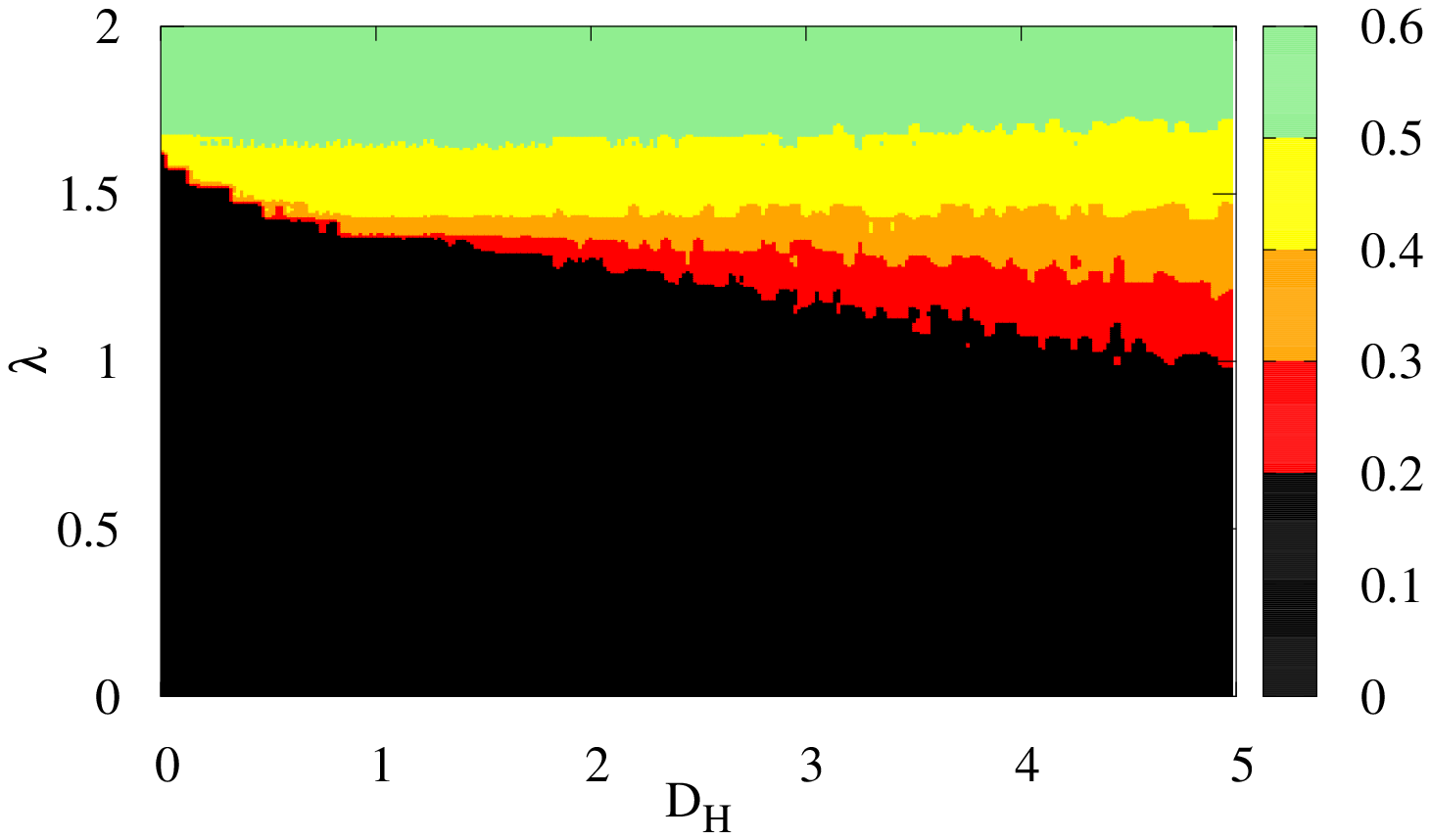}}
 \caption{Parameter space diagram in $D_H-\lambda-R$ space with (a) triangular and (b) uniform frequency distribution for the system Eq.~\eqref{6}. Color axis represents the value of $R_{amp}$.}
 \label{fig10}
 \end{figure}

Here we study the variation of the amplitude order parameter $R_{amp}$ with the noise strength in hub $(D_H)$ for different values of coupling strengths. Parameter space diagram of the system Eq.~\eqref{6} with triangular and uniform FD is shown in Figs.~\ref{fig10}(a) and \ref{fig10}(b) respectively. Simulations are done using the uniformly distributed phases between [0:2$\pi$] as initial conditions for each $D_H$-value and then using the successive state as the initial condition for next $\lambda$. In the absence of noise, the system is desynchronized till the forward transition point. If the noise is switched on, the system can attain synchrony even below the forward transition point. For smaller values of $D_H$ the transition remains first order for both the triangular and uniform FDs as shown in Figs.~\ref{fig10}(a) and \ref{fig10}(b) respectively. As we increase the value of $D_H$, the order parameter changes continuously indicating that the transition is a second order transition. Thus, we can see that noise induces partial synchrony in the system for the parameter values where the system would otherwise behave incoherently.

\section{summary}
\label{sec4}
In this paper we have studied the transition to explosive synchronization in the presence of noise. The dynamics on each node is either a Kuramoto oscillator or a Stuart-Landau oscillator. We have considered a case where noise is present in hub. We observe that the critical value of forward transition point in explosive synchronization and the width of hysteresis area crucially depends on the noise strength in the hub $(D_H)$. If the noise is present in the hub, we observe a shift in the forward transition point while there is no change in the backward transition point with increase in the noise strength ($D_H$). The shift in the forward transition point results in a decrease in the hysteresis area with increasing $D_H$. At a certain value of noise strength, the region of hysteresis width is minimum after which the transition changes to a second order transition. The second order transition to synchrony takes place at coupling values $\lambda<\lambda_b^c$ (backward critical transition point).

We have considered model example of Kuramoto and SL oscillators in presence of noise. We observe that presence of noise in hub may change the hysteresis area or the nature of transition may become continuous. When the strength of noise in the hub is small, the hysteresis width may decrease. Depending upon the noise strength, the hysteresis loop may vanish after which the transition becomes a second order transition. We have also calculated the transition points for Kuramoto oscillators using semi-analytical approach and observe that it is in good agreement with the numerical findings.
Thus, this scheme provides an effective way to control explosive synchronization by tuning noise strength in the hub. 

Our results are helpful in testing the robustness of explosive synchronization observed in case of star networks. 
This study can further be extended by exploring explosive synchronization in more complex physical, biological systems or chaotic systems. Further, the presence of noise may provide an additional mean to induce explosive synchronization in the system.

\section*{ACKNOWLEDGMENTS}
RV wants to acknowledge the financial assistance given by CSIR, India under the file no.09/112(0601)/2018-EMR-I. HHJ would like to thank UGC, India for the award of grant no. F:30-90/2015 (BSR).  We also thank R. Ramaswamy for useful discussions.


\begin{thebibliography}{1}
 
\bibitem{Pikovsky-Book} A. Pikovsky, M. Rosenblum, and J. Kurths, {\it Synchronization: A Universal Concept in Nonlinear Science} (Cambridge University Press, Cambridge, 2001).

\bibitem{Boccaletti-Book} S. Boccaletti, A. N. Pisarchik, C. I. D.  Genio, and A.  Amann, {\it Synchronization: From Coupled Systems to Complex Networks} (Cambridge University Press, Cambridge, 2018).

\bibitem{abrams}D. M. Abrams and S. H. Strogatz,\prl{93} {174102} {2004}.

\bibitem{Strogatz-Book} S. H. Strogatz, {\it Sync: The Emerging Science of Spontaneous order} (Hyperion, New York, 2003).

\bibitem{Arenas-2006}A. Arenas, A. Diaz-Guilera, and C. J. Perez-Vicente, \prl{96}{114102}{2006}.
 
 \bibitem{Arenas-2008}A. Arenas, A. Diaz-Guilera, J. Kurths, Y. Moreno and C. Zhou,  Physics Reports {\bf 469}, 93 (2008).
 
 \bibitem{Barahona-2002}M. Barahona and L. M. Pecora, \prl{89}{054101}{2002}.
 
\bibitem{watts}D.J. Watts, S.H. Strogatz, Nature {\bf 393} (1998) 440.

 \bibitem{Gardenes-2007}J. G\'omez-Garde\~nes, Y. Moreno, and A. Arenas,\prl{98}{034101}{2007}.
 
 \bibitem{Nishikawa-2003}T. Nishikawa, A. E. Motter, Y. C. Lai, and F. C. Hoppensteadt, \prl{91}{014101}{2003}.
 \bibitem{Kuramoto-Book} Y. Kuramoto, {\it Chemical Oscillations, Waves, and Turbulence}, (Springer, New York,1984).
 
 \bibitem{Acebron-2005}J. A. Acebr\`on, L. L. Bonilla, C. J. P. Vicente, F. Ritort, and R. Spigler, \rmp{77}{137}{2005}.
 
 \bibitem{Gardenes-2011} J. G\'omez-Garde\~nes, S. G\'omez, A. Arenas, and Y. Moreno, \prl{106}{128701}{2011}.


\bibitem{Peron-2012} T. K. D. M. Peron, and F. A. Rodrigues \pre{86}{016102}{2012}.
%
\bibitem{Zhang-2013} X. Zhang, Xin Hu, J. Kurths, and Zonghua Liu, \pre{88}{010802}{2013}.
%
\bibitem{Hu-2014} X. Hu, S. Boccaletti, W. Huang, X. Zhang, Z. Liu, S. Guan, and C. H. Lai, \screp{4}{7262}{2014}.
%
\bibitem{Zhou-2015}W. Zhou, L. Chen, H. Bi, X. Hu, Z. Liu and S. Guan,\pre{92}{012812}{2015}.
%
\bibitem{Coutinho-2013} B. C. Coutinho, A. V. Goltsev, S. N. Dorogovtsev, and J. F. F. Mendes, \pre{87}{032106}{2013}.

\bibitem{Zou-2014} Y. Zou, T. Pereira, M. Small, Z. Liu, and J. Kurths, \prl{112}{114102}{2014}.

\bibitem{Vlasov-2015}V. Vlasov, Y. Zou, and T. Pereira, \pre{92}{012904}{2015}.

\bibitem{Yeung1999} M. K. Stephen Yeung and Steven H. Strogatz, \prl{82}{648}{1999}.

\bibitem{Bi-2014}H. Bi, Xin Hu, X. Zhang, Y. Zou, Z. Liu and S. Guan, \epl{108}{50003}{2014}.
%
\bibitem{Levya-2012}I. Leyva, R. Sevilla-Escoboza, J. M. Buldú, I. Sendiña Nadal, J. Gómez-Gardeñes, A. Arenas, Y. Moreno, S. Gómez, R. Jaimes-Reátegui, and S. Boccaletti, \prl{108}{168702}{2012}.

\bibitem{Chen-2013} H. Chen, G. He, F. Huang, C. Shen, and Z. Hou, \chaos{23}{033124}{2013}.

\bibitem{Boaretto-2019} B. R. R. Boaretto, R. C. Budzinski, T. L. Prado, and S. R. Lopes \pre{100}{052301}{2019}.

\bibitem{Ji-2013} P. Ji, T. K. DM. Peron, P. J. Menck, F. A. Rodrigues, and J. Kurths, \prl{110}{218701}{2013}.
%
\bibitem{Ji-2014}P. Ji, T. K. DM. Peron, F. A. Rodrigues, and J. Kurths, \pre{90}{062810}{2014}.


%
\bibitem{Filatrella-2007} G. Filatrella, N. F. Pederson, and K. Wiesenfeld, \pre{75}{017201}{2007}.
%
\bibitem{Kumar-2020} A. Kumar , S. Jalan, and A. D. Kachhvah, \prr{2}{023259}{2020}.
\bibitem{Jalan-2019}S. Jalan, V. Rathore, A. D. Kachhvah, and A. Yadav, \pre{99}{062305}{2019}.
\bibitem{Kumar-2021}A. Kumar and S. Jalan, \chaos{31}{041103}{2021}.

\bibitem{Ojalvo-1996} J. Garc\'ia-Ojalvo, R. Roy, \pla{224}{51}{1996}.

\bibitem{Greenman-2003} J.V. Greenman, T.G. Benton, Amer. Nat. {\bf 161}, 225 (2003).

\bibitem{Chen-2004} D . Chen, M.A. Cane, A. Kaplan, S.E. Zabiak, D. Huang, Nature {\bf 428} {733}(2004).

\bibitem{Surovyatkina-2005}E. Surovyatkina, Nonlinear Process. Geophys. {\bf 12}, 25 (2005).

\bibitem{Alonso-2007} D. Alonso, A.J. McKane, M. Pascual, J. R. Soc. Interface {\bf 4} 575 (2007).

\bibitem{Erguler-2008} K. Erguler, M.P.H. Strumpf, Math. Biosci. {\bf 216}, 90 (2008).

\bibitem{Benzi-1981} R. Benzi, A. Sutera, and A. Vulpiani, \jpa{\bf 14}{453}{1981}.
\bibitem{Longtin-1998}A. Longtin, and D. R. Chialvo, \prl{81}{4012}{1998}.
\bibitem{Kaneko-1997}K. Kaneko, \prl{78}{2736}{1997}.

\bibitem{Kraut-1999}S. Kraut, U. Feudel, C. Grebogi, \pre{59}{5253}{1999}.

  \bibitem{Pisarchik-2009} A.N. Pisarchik, R. Jaimes-Re\'ategui, \pla{374}{228}{2009}.

\bibitem{Zerega-2012}B.E. Mart\'inez-Z\'erega, A.N. Pisarchik, \cns {\bf 17}{4023}{2012}.


\bibitem{Glass-2001} L. Glass, Nature {\bf 410}, 277 (2001).
\bibitem{Lloyd-1999}A. L. Lloyd and R. M. May, Trends Ecol. Evol. {\bf 14}, 417 (1999).
\bibitem{Neiman-1999}A. Neiman, X. Pei, D. Russell, W. Wojtenek, L. Wilkens, F. Moss, H. A. Braun, M.
T. Huber, and K. Voigt, \prl{82}{660}{1999}.
\bibitem{Tavazoie-1999}S. Tavazoie, J. D. Hughes, M. J. Campbell, R. J. Cho, and G. M. Church, Nat.
Genet. {\bf 22}, 281 (1999).
\bibitem{Tu-2005} P. Tu, A. Kudlicki, M. Rowicka, and S. L. McKnight, Science {\bf 310}, 1152 (2005).
%\bibitem{Skardal-2014} P. S. Skardal and A. Arenas, \pre{89}{062811}{2014}.
\bibitem{Uchida-2004}A. Uchida, R. McAllister, and R. Roy, \prl{93}{244102}{2004}; J. N. Teramae and D. Tanaka, \prl{93}{204103}{2004}.
\bibitem{Nagai-2010}Ken H. Nagai, and Hiroshi Kori, \pre{81}{065202(R)}{2010}.

\bibitem{Sakaguchi-1998} H. Sakaguchi, Prog. Theor. Phys. {\bf 79}, 39(1988); S. H. Strogatz and R. E. Mirollo, \jsp{\bf 63}{613}{1991}.

\bibitem{Cao-2018} L. Cao, C. Tian, Z. Wang, X. Zhang, and Z. Liu, \pre{97}{022220}{2018}.

\bibitem{Watanabe-1993} S. Watanabe and S. H. Strogatz, \prl{70}{2391}{1993}.

\bibitem{Watanabe-1994} S. Watanabe and S. H. Strogatz, \physd{74}{197}{1994}.

\bibitem{Marvel-2009} S. A. Marvel, R. E. Mirollo, and S. H. Strogatz, \chaos{19}{043104}{2009}.
\bibitem{Pikovsky-2015} A. Pikovsky and M. Rosenblum, \chaos{25}{097616}{2015}.

 
\bibitem{beta} S. Kotz and J. R. V. Dorp, {\it Beyond Beta: Other Continuous Families of Distributions with Bounded Support and Applications} (World Scientific, 2004).

\bibitem{Rosenblum-1996} M. G. Rosenblum, A. S. Pikovsky, and J. Kurths, \prl{76}{1804}{1996}.

\bibitem{Matthews-1991}Paul C. Matthews, R. E. Mirollo and S. H. Strogatz, \physd{52}{293}{1991}.

\end{thebibliography}
\end{document}